\documentclass[10pt,conference]{IEEEtran}
\IEEEoverridecommandlockouts
\usepackage{cite}
\usepackage[numbers]{natbib}
\usepackage{amsmath,amssymb,amsfonts}
\usepackage{algorithmic}
\usepackage{comment}
\usepackage{graphicx}
\usepackage{textcomp}
\usepackage{xcolor}
\def\BibTeX{{\rm B\kern-.05em{\sc i\kern-.025em b}\kern-.08em
    T\kern-.1667em\lower.7ex\hbox{E}\kern-.125emX}}

\usepackage{listings, xcolor}
\usepackage{graphicx}
\usepackage{amssymb}

\usepackage{lineno}
\usepackage[numbers]{natbib}

\usepackage{blindtext}

\usepackage{url} 
\usepackage{hyperref}
\begin{document}

\title{Automatic Generation of Blockchain Agri-food Traceability Systems}

\author{\IEEEauthorblockN{Lodovica Marchesi}
\IEEEauthorblockA{\textit{Dept. of Mathematics}\\
\textit{and Computer Science} \\
\textit{University of Cagliari, Italy}\\
lodovica.marchesi@unica.it}
\and
\IEEEauthorblockN{Katiuscia Mannaro}
\IEEEauthorblockA{\textit{Dept. of Mathematics}\\
\textit{and Computer Science} \\
\textit{University of Cagliari, Italy}\\
katiuscia.mannaro@unica.it}
\and
\IEEEauthorblockN{Raffaele Porcu}
\IEEEauthorblockA{\textit{Dept. of Mathematics}\\
\textit{and Computer Science} \\
\textit{University of Cagliari, Italy}\\
raffaele.porcu@unica.it}
}

\maketitle

\begin{abstract}
Supply chain management, product provenance and quality certification are among the first and most popular applications of blockchain technology, due to the inherent trust and inalterability provided by the blockchain technology. However, the proposed supply chain management systems based on blockchain and smart contract technology tend to be specific to the particular production and production process. In this paper we present a general-purpose approach for the agri-food supply chain management, proposing a system that can be configured for most agri-food productions. The primary purpose is to provide a methodology to facilitate and make more efficient the development of such applications. It is based on general smart contracts and apps interacting with the same smart contracts, which are configured, starting from the description of the specific system to be managed, given using json files. A case study on olive oil production is described, to show how our approach works.
\end{abstract}


\section{Introduction}

Food fraud has become a  issue on a global scale for producers, consumers, governments and other involved actors \cite{Galvez2018}. In fact although it is not a new problem - fraud and adulteration of food were widespread during ancient times so much so Roman laws included edicts against the commercial frauds or adulteration - unfortunately appears rising in the last years and is  a threat to public health and the economy.

According to \citet{spink2011defining}  \emph{``Food fraud is a collective term used to encompass the deliberate and intentional substitution, addition, tampering, or misrepresentation of food, food ingredients, or food packaging; or false or misleading statements made about a product, for economic gain.''}


The globalization of the markets reduces the food traceability and so facilitates food fraud. Consequently, food safety can be increased through higher traceability \cite{Galvez2018}. 
In order to identify and address food frauds worldwide, it is essential to trace and authenticate the food supply chain to understand provenance and to ensure food quality in compliance with international standards and national legislation. 
In the literature several definitions of traceability and of traceability system can be found  \cite{Olsen2013}, \cite{Bosona2013}, \cite{Galli2015}, \cite{Karlsen2013}, \cite{Dabbene2014}.
Food traceability is \emph{``the ability to follow the movement of a feed or food through specified stage(s) of production, processing and
distribution''} \cite{ISO22005_2007}  as defined by  the International Organization for Standardization in ISO 22005:2007 - a specific standard for traceability in the food and feed chain. In addiction, a traceability system is based on product labelling and contains information, both quantitative and qualitative, about the final product and its provenance.
In Europe, food legislation is particularly strict and the implementation of traceability systems are mandatory but are unable to insure  consumers against fraud. For this reason innovative methods for traceability systems based on product identification are needed.
 Deterring food fraud requires interdisciplinary research combining food science, food law, supply chain management, and other fields such as informatics, mathematics and statistics \cite{DANEZIS2016123}.

In this context, a distributed ledger technology such as Blockchain (BC) provides a full and immutable audit trail of transactions data for all stages of the food supply chain  allowing for transparency and verifiable and immutable records in the form of digital certificates.

Immutability of the data enables the technology to be considered for regulated industries such as agri-food.  

As reported by \citet{Galvez2018} in his research paper most blockchain systems for traceability management have been developed since 2015. 

Numerous research papers, \citet{Demestichas2020}, \citet{costa2013review}, \citet{tian2016agri}, \citet{Antonucci2019} just to cite a few,  have shown that using blockchain can advantageously help to achieve traceability by irreversibly and immutably storing data. 
Although many studies are promising that the application of blockchain technologies to the agri-food supply chain can ensure the food traceability and some companies have launched pilot or proof of concept projects to manage their supply chains with blockchain technology, certain limits remain to be considered and addressed. Most of the published works concerning  the application of blockchain technologies to the supply chain management have reported no detailed information about the technical implementation and there are still few uses to the state of practice (See Section \ref{sec:relatedWork}).

For this reason we investigated the potential applications of blockchain technologies in the agri-food sector both from some large companies that have declared their intention to develop pilot or proof of concept projects and several research papers in literature  about the possibility of applying such technologies in agri-food domain. Our investigations indicated us a main direction of research: devising new methods for agri-food supply chain implementation processes based on blockchain technology. In our opinion research in this topic area should investigate how blockchain-based processes for implementing the agri-food supply chain traceability can be efficiently specified and deployed. Among others, formal methods  to facilitate and make more efficient the development of applications on blockchain technologies for the management of the agri-food supply chain. Researchers and developers could  benefit from a general-purpose approach for agri-food supply-chain management.



In light of these considerations and to better understand this technology and generate new user-friendly implementations, the primary purpose of this paper is to propose a blockchain-based methodology to facilitate and make more efficient the development of blockchain applications for the agri-food  supply chain management, by building configurable blocks to be assembled together, so as not to start from scratch every time. Finally, we evaluated the proposed approach using a specific case study: a traceability system for the olive oil supply chain.

This methodology can help to compose Smart Contracts (SCs) specific in an easy way, so as to be able to reuse the code and modules and automate the process to make it fast, keeping its safety. 
As far as we know, this is the first attempt to develop a semi-automatic configurable system that supports the entire class of supply chains for agri-food industrial domain.
The rest of the paper is organized as follows. Section \ref{sec:relatedWork} deals with the background and related work. Section \ref{sec:Methodology} presents the proposed methodology, followed by Section \ref{sec:Case study} including a case study on a traceability system for the olive oil supply chain to evaluate the proposed methodology. Section \ref{sec:Conclusions} makes a brief conclusion for this paper.

\section{Background and Related Work} \label{sec:relatedWork}

In this section, we first introduce an overview on decentralized application and smart contracts. Then, we briefly introduce some background knowledge of related work which leads the readers to better understand the issue.

A decentralized application, or DApp, is a computer application that runs on a distributed system, that is on a network of nodes, with no node acting as supervisor. In the context of blockchain peer-to-peer network, a DApp is stored and executed on the blockchain, in order to be decentralized, transparent, deterministic and redundant. A DApp is developed by writing Smart Contracts (SCs), which are small script programs running on every node of the blockchain and may have a User Interface (UI) that allows users and devices to interact with them.
SCs are immutable - no one can tamper with the code of the contract - and distributed, because of their storage inside the blockchain
Smart contracts have been developed and used in several fields \cite{zheng2018blockchain}; here we focus on agri-food industrial supply chains.

Based on various definitions of traceability systems (\cite{Olsen2013}, \cite{Bosona2013}, \cite{Galli2015}, \cite{Karlsen2013}, \cite{Dabbene2014}, \cite{ISO22005_2007}, just to cite a few),  the main purposes of a agri-food traceability system are:
\begin{itemize}
    
\item To document in a transparent and irreversible way any event relevant to production;
\item To allow authorities, laboratories and certified experts to asseverate the production, giving proof of their identity and their certifications;
\item To integrate manual registrations and automatic  registrations, made by Internet of Things (IoT) devices that are increasingly widespread;
\item To keep track of the quantities produced, so that these cannot be increased by introducing products of non-certified origin;
\item To give evidence of all stages of production to the authorities responsible for verifying the specifications;
\item To allow retailers and end consumers to learn about the history of the products purchased, from the field to the purchased product, using an app.

\end{itemize}

Traditional approaches for the agri-food traceability systems  are usually inefficient. Blockchain is a promising technology that is tamper-proof and decentralized and  self-executing and self-verifying smart contracts are able to conduct transactions between mutually untrusted parties. In this context scholarly literature on the adoption of the blockchain technology in specific traceability systems for the agri-food supply chain is beginning to emerge. 
In particular, since 2018, a lot of research efforts have been made on the use of blockchain technology for traceability systems. \cite{Demestichas2020}
 
  Since 2018, Tripoli and Schmidhuber highlighted the potential of distributed systems to transform the agri-food industry \cite{tripoli2018emerging}. \citet{tribis2018supply} searched for relevant papers about the adoption of blockchain technology for generic supply chain management, covering the literature from all years to 2018, and found out that most of the papers were focused on the use of BC for traceability, but only one out of 40 papers dealt with agricultural field. \citet{chang2020blockchain} claim that by 2023 the global blockchain supply chain market will grow to \$ 3,314.6 million, recording an increase in the annual growth rate of 87\% . 
Various papers presented real blockchain solutions for supply chain management, which proved to be successful. Among others, AgriDigital \footnote{https://www.agridigital.io/}, an Australian system for managing grain supply chain released in 2017, counts at the end of 2020, more than 7000 users and a transaction value of \$ 3,793 million.
\citet{caro2018blockchain} developed AgriBlockIoT, a fully decentralized, blockchain-based traceability solution for agri-food supply chain management, able to seamless integrate IoT devices, using and comparing both Ethereum and Hyperledger Sawtooth blockchains. \citet{tian2016agri} studied and developed an agri-food supply chain traceability system for China based on RFID (Radio-Frequency Identification) and blockchain technology, to guarantee food safety. \citet{baralla2019ensure} proposed a generic agri-food supply chain traceability system based on blockchain technology implementing the "from-farm-to-fork" (F2F) model currently used in the European Union, which can integrate current traceability rules and processes, using Hyperledger Sawtooth, and following an agile approach.


\citet{wang2019smart} propose a product traceability system based on blockchain technology, in which all product registration and transfer histories are perpetually recorded by using smart contracts and an event response mechanism was designed to verify the identities of both parties of the transaction and guarantee the validity of the transaction.

\citet{Yu2020a} propose a monitoring framework that combines smart contracts and evaluation models for the automatic evaluation of the quality of fruit juice samples. Smart contracts are executed to record production data on a blockchain and can decide whether the production process can be resumed or not. The feasibility of the system has been evaluated by implementing a prototype version of the quality monitoring system for flat peach juice production based on the Ethereum platform and executed in the Remix IDE.

Many studies in the literature have focused on highlighting the benefits and value derived by the  blockchain implementation in agri-food supply chain domain. 
In particular, \citet{kamble2020modeling} conducted a analysis in order to model a traceability system based on blockchain technologies in the agriculture supply chain.  They identified  thirteen enablers that encourage the blockchain adoption in agriculture supply chain,  such as anonymity and privacy, immutability, smart contracts, secured and shared database, traceability, transparency, and others. Then they established hierarchical levels and relationships between the involved actors in the supply chain through interpretive structural modeling (ISM) and decision-making trial and evaluation laboratory (DEMATEL) methodologies.  The enablers were identified from existing literature and validated by experts from the field of agrobased supply chain and technology.
Moreover the authors conducted an interesting literature review revealing that BC technology offers various benefits leading to improvements in the sustainable performance of the agriculture supply chains. 

Based on previous research papers (\cite{tian2016agri}, \cite{tian2017supply}, \cite{kamble2020modeling}, \cite{wang2019understanding}, \cite{Li2018}), just to cite a few, by using a blockchain the traceability system guarantees:

\begin{itemize}

\item Data integrity and provenance of documents and records on blockchain;
\item Immutability and transparency of data recorded on the blockchain; 
\item Respect for the quantities of the products involved (grapes, wine, bottles), based on the annual production of the land and the yield in the various stages of processing. This is achieved with the system of tokens, which are associated with the various products and cannot be altered as they are managed on the blockchain;
\item Buyers take charge of their products, with complete traceability of the supply chain;
\item Ability to retrace the entire supply chain, simply by accessing the blockchain and public servers with documents, starting from the QR code shown on the final product.
\end{itemize}

To the best of our knowledge, and by comparing our work to the others that deal with the use of DApps and blockchain to certify the origin of food and prevent food fraud, we found out that this is the first paper that proposes the use of a self-generating system.

\section{Methodology} \label{sec:Methodology}
We studied the agri-food production domain, and we found that virtually all agri-food supply chain systems start from one or more primary sources (soil, herd, beehives, lake, sea, and so on.), produce a primitive resource (harvest, milk, raw honey, fishes, etc.), then this product is transformed possibly several times, until the final product; events relevant to the process can occur in each of the phases.
This problem is too complex to be tackled top-down, therefore we adopted a bottom-up approach. Using analysis techniques and object-oriented design, we performed the analysis of the supply chain domain in order to:

\begin{itemize}

 \item Determine the actors, concepts and entities that emerge and recur in this type of systems;
 \item Determine the relationships between these entities;
 \item Determine the events of interest for tracking and for the generation of new products (transformations, splitting, merging);
 \item Design configurable and modular SCs/building blocks, representing the identified entities and relationships, which can then be represented on the blockchain;
 \item Semi-automatically map the user interface needed to interact with the system, describing the entities, events and transformations.
 \end{itemize}
 
In this section we present our approach and we show how the traceability system is then generated automatically, saving time and costs. The entire software process was carried on following an agile approach such as ABCDE (Agile BlockChain Dapp Engineering) \cite{marchesi2020abcde}.

\subsection{Actors}

We define the key actors of an agri-food system. Each actor is identified by a unique address able to send transactions to the blockchain. The actor owns the private key associated with the address, thus being the only person able to send messages from that address.
\begin{itemize}
   
\item  \textbf{Administrator/Owner}: Administers the system, managing the addresses of the other actors and their permissions.
\item \textbf{Producer}: Produces the raw materials that are the inputs of the supply chain.
\item \textbf{Supplier}: Supplies materials, services or devices needed for producing the target good.
 \item \textbf{Transformer}: Transforms raw material, or already transformed material, into intermediate goods or into the target good.
\item \textbf{Wholesaler}: Buys the target good in large quantities, and sell it to other wholesalers or retailers.
\item \textbf{Retailer}: Buys the target good in order to sell it to End Customers.
\item \textbf{End Customer}: Buys the target good from Retailers. S/he is usually not provided with an address.
\item \textbf{Certification Authority}: Public or private authority in charge of controlling and certifying a given production.
\item \textbf{Professional}: A person with a given degree and experience, qualified to certify the goodness of amounts and documents.
\item \textbf{Analysis Lab}: A laboratory able to perform physical, chemical and/or biological analysis of given materials, products or goods.
\item \textbf{Warehouse}: Receives, stores and sends goods.
\item \textbf{Device}: An IoT device connected to the Web, able to send transactions with measurements relevant to the supply chain.
\end{itemize}


\subsection{Entities}

The main entities involved in an agri-food supply chain are:

\begin{itemize}
\item \textbf{Address Catalog}: for each address, the identity and the role(s) of the owner are specified; the catalog is managed by the Administrator/Owner of the system.
\item \textbf{Producer}: a farmer or a firm producing or transforming agri-food products. Its representative(s) are identified by blockchain addresses.
\item \textbf{Productive Resource}: it represents something that produces the main raw agri-food products. Typically, it is a field (orchard, vineyard, wheat field, vegetable garden, greenhouse, olive grove, etc.), a group of animals (flock, herd, poultry, etc.), a set of beehives. It is owned by a Producer. The system can hold information and documents on it, and can register events related to its cultivation or farming.
\item \textbf{Product}: it represents a production lot of something that comes from a Productive Resource, or from the transformation of other products. For instance, the grapes are produced from the vineyard, the must from the pressed grapes, the wine from the must.  It is linked to a Producer which is in charge of its processing. The system can hold information and documents on it, and can register events related to its processing.
\item \textbf{Token}: a given quantity (a number) created and assigned to a given address. The token represents the ownership of a specific amount of material, good or asset. It can be split, and transferred to other address(es). Since the token, once created, cannot be increased, it guarantees that only the original material/good/asset is managed in the system. Many kinds of tokens can be managed by the system.
\item \textbf{Notarization} of documents assessing the process (treatments, harvest, chemical analysis, quantity produced in subsequent steps, etc.). This is a known pattern [10]. The original document must be kept on a server, and available to download to authorized users. The notarization must include:

    \begin{itemize}
    \item hash of the document;
    \item registration date (always available as date and time of the transaction);
    \item link to the document in the server, or information on how to access it;
    \item possible metadata.
    \end{itemize}
\end{itemize}

\subsection{Main Events}
There are two types of events:

\begin{enumerate}
    \item \textbf{Transformation events}, which create a product starting from one or more resources, transform one or more products into one or more others, or divide the product into several sub-products that are the same, but of lower quantities;
    \item \textbf{Documentation events}, which associate data related to the production process to a product (or resource), but do not transform it and do not create other products.
\end{enumerate}

The main events managed by the system are:
\begin{itemize}
 
\item \textbf{Asseveration}: a given entity stored on the blockchain (hash of a document, data, etc.) is certified by a transaction sent from an address of a person/body able to certify. Also the correctness of product and token creation or transformation can be asseverated.
\item \textbf{Creation of tokens} associated to some product of the process, at a given date. The creation can be provided of further data about the physical product associated to the tokens, and even of the notarization of documents attesting the truthfulness of the creation.
\item \textbf{Product Merging}: the act of merging products of the same kind, invalidating them and producing products of the same kind. Different lots of the same material can be merged, producing one or more new lots. The tokens associated to the products must be burned, and new tokens associated to the new products are created, preserving the overall number of tokens.
\item \textbf{Product Splitting}: the act of splitting one product into two or more of the same kind, invalidating the original and producing products of the same kind. Also in this case the tokens associated to the original product must be burned, and new tokens associated to the new products are created, preserving the overall number of tokens.
\item \textbf{Product Transformation}: the act of merging one or more products into one or more products of different kinds, invalidating the original one(s) and producing products of one or more other kinds. For instance, grapes can be transformed to must, used to produce wine, but also to marc, used to produce grape pomace brandy. Also in this case the tokens associated to the original products must be burned, and new tokens associated to the new products are created, preserving the overall balance of quantities. The transformation is often associated to asseveration events.
\item \textbf{Data Registration}: a record holding specific data is stored in the blockchain by a given address, guaranteeing the date and the actor who stored it. For instance, fertilization, pesticide treatments and pruning are recorded as events linked to a field (Productive Resource).
\item \textbf{Notarization Event}: the data concerning a Notarization (see previous section) are registered by a specific address, ensuring date and inalterability of the documentation, and signature of the registrant.
\item \textbf{Certification of data or of a notarization}: a third party certifies, with a transaction coming from its address, the correctness of a data registration, or of a notarized document.
\item \textbf{Unlocking}: one or more transactions from given addresses are needed to unlock a process, that is to register another event (typically, a transformation event).
\item \textbf{Payment}: a payment (in cryptocurrency) is made available to a given address.
\end{itemize}

\subsection{Smart Contracts}
After identifying the actors, concepts and events related to the process, the system SCs are created.
Depending on the specific use case to be implemented, the
developer can design either dynamic, static, or oracle driven
smart contracts. 
At the moment in our system the smart contract is a static file regardless of the use case (olive or wine or honey supply chain use case, just to cite a few)  listing all the relevant fields for each concept, as described in the following section. The SCs are automatically generated from the JSON file. We will have a SC for each producer (\texttt{Producer}), and a SC for each resource (\texttt{ProductiveResource}) and for each product being processed or transformed (\texttt{AgriProduct}), which takes into account the registrations (\texttt{AgriEvent}) on it. Since a \texttt{ProductiveResource} and an \texttt{AgriProduct} share several data and operations, both inherit from \texttt{AbstractResource} abstract SC.
The \texttt{Producer} contains a mapping with links to all productive resources and products owned by, or related to it. Products are generated or transformed starting from one or more productive resources, or from one or more products. Each product tracks its origin or origins through an array of addresses of the upstream resources or products. An \texttt{AbstractResource} holds a list of events \texttt{AgriEvent}, which in turn contain a list of data related to the event (all encoded in the string of bytes named "parameters"), to flexibly attribute data to the events, as per the above approach for the GUI. 
A QR code on the final product allows to find the related \texttt{AgriProduct}. From here you can find all the events in (reverse) registration order and, for each event, all relevant data, including links to documents and their possible hash digest. If the \texttt{AgriProduct} has one or more origins, by going to these further origins (\texttt{AgriProduct} or \texttt{ProductiveResource}), all upstream products/resources can be found, with the related events, and so on. At the end of each product, you can also access its \texttt{Producer}. 
Backward navigation must be thoroughly designed, to make it easier for multiple origins. If necessary, a list of generic data could also be added to the \texttt{AbstractResource}, encoded in a string of bytes, as in the "parameters" field of \texttt{AgriEvent}. The relationship between product and upstream products/resources is navigable in both directions. An \texttt{AgriProduct} contains the array "origins", with the addresses of the contracts of type \texttt{ProductiveResource} or \texttt{AgriProduct} that created it (the \texttt{ProductiveResource} has no origins). This is a list created during the creation of the \texttt{AgriProduct}, and cannot be modified. A generic product/resource also contains a “produced” array with the addresses of the generated \texttt{AgriProduct}, for Creation, Transformation, Division or Contribution. This array can be updated, typically by appending the address of a newly created \texttt{AgriProduct}. 
It is recommended to refer to some best practices given by researchers or communities also to avoid vulnerabilities.
In order to optimize the code, we use OpenZeppelin \cite{OpenZeppelin} - a toolkit to develop, compile, update, deploy and interact with Smart Contracts. We also systematically applied gas saving patterns \cite{marchesi2018gas}.

\begin{figure}[h]
\centering\includegraphics[width=1\linewidth]{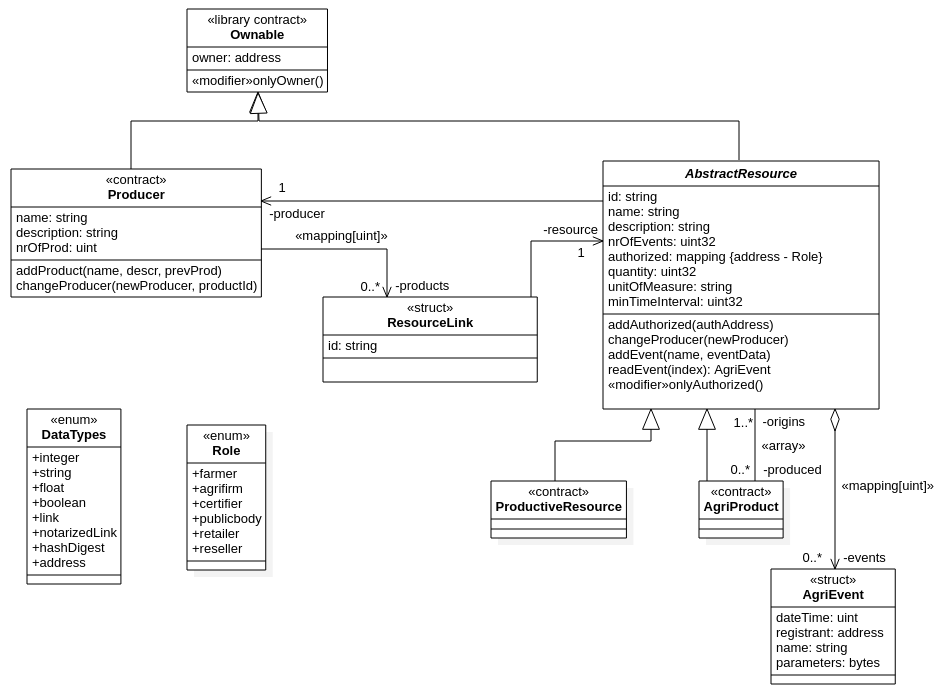}
\caption{UML class diagram representing the Smart Contracts used in our system.}\label{fig:UML}
\end{figure}

Figure \ref{fig:UML} shows the UML class diagram representing the Smart Contracts used in our system.

\subsection{Data Types}
All product and events are characterized by data which are the same in every agri-food supply chain, such as type, name, unique identifier, owner. However, they may hold many more data, specific of the particular production, and production process.

Since our goal is to develop a general-purpose system, able to be configured for every agri-food production process, we use a flexible data representation. Besides the above quoted common data, each element can be provided of a list of data, each   represented by a string. The string representing a data includes three substrings giving the name of the data, its type, chosen among a set of common types, and its value. The types allowed by the system are: int, float, string, text (multiline string), enum, link, hashlink (link to an hashed file, stored off-chain), upload (it is possible to upload a local file or a photo to a server, keeping its link on the blockchain), hashupload (same of upload but the file is firstly hashed, and its hash digest is stored in the blockchain).
For instance, the harvest event could have associated the list of parameters that describe how many kilograms were collected (type int), the grape variety (type string) and even a photo taken during the harvest (type hashupload).

\subsection{User Interface}

\begin{figure}[h]
\centering\includegraphics[width=1\linewidth]{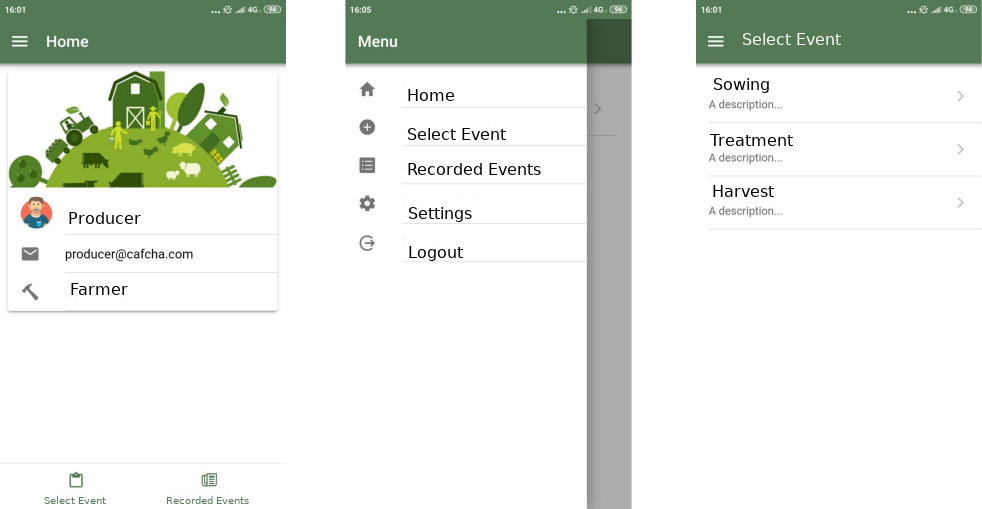}
\caption{Homepage of an operator, menu and selection of the recordable events.}\label{fig:HomePageOperator}
\end{figure}

\begin{figure}[h]
\centering\includegraphics[width=1\linewidth]{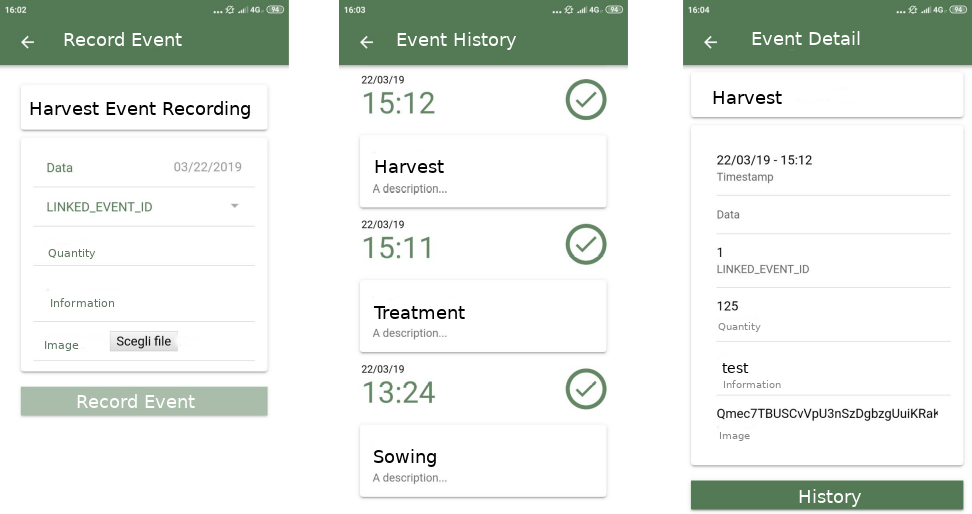}
\caption{Recording an event and navigating a product's event history.}\label{fig:NavigazioneEventi}
\end{figure}

The entities and the events of an agri-food supply chain, as defined in the sections above, are standard, and are able to cover virtually every production process. Consequently, also the applications enabling their input, editing (when allowed), and retrieving, are standard, and their user interface (UI) can be automatically generated starting from the description of the system.
More precisely, once the events of a specific production process are defined, with their data, constraints and authorizations, it is possible to automatically generate an app (a HTML5 Web page provided of Javascript code) able to create and edit the event. Clearly, the style and the appearance of the UI can be customized, but the data input, with all proper checks, do not require further programmer's intervention.
Also the navigation among the events, the products and the productive resources can be automatically programmed, starting from a QR code written on the final product. This navigation does not require a blockchain address, and can be performed by every customer.
Figures \ref{fig:HomePageOperator} and  and \ref{fig:NavigazioneEventi} show some possible screenshots of the application, automatically generated.

\section{The Case Study: a blockchain traceability system for the olive oil supply chain} \label{sec:Case study}

The object of the case study is a blockchain traceability system to trace the olive oil production, from the agricultural treatments, to the olive harvest, to the oil mill, to the bottled oil, until the sale of bottles to a wholesaler and eventually to shops and supermarkets.
The choice is not accidental because olive oil and in particular Extra-Virgin olive oil remains today one of the most fraudulent food products.
The architecture of the system includes a cloud space as off-chain data repository. In this case, it has been used a decentralized technology, with Ipfs, an App to send, register and read events (written in Javascript using Angular/Ionic framework), the use of a real blockchain, such as Ethereum testnet, with app to send transactions or read from blockchain (Javascript, web3.js, and so on) and finally Smart Contracts written in Solidity.

A well-known issue with DApps developed on Ethereum-like blockchains is the cost of gas, which is specially related to the amount of Storage memory used. Solidity provides different types of memories. Those of interest to our application are:
\begin{itemize}
    \item Storage: it is the persistent memory, and the most expensive;
    \item Event-log: it stores on the blockchain data related to the Events raised by SCs (these Events are the syntactic constructs of Solidity). These data are cheap, but can be accessed only by external applications.
\end{itemize}

The information related to actors, entities and events may be too big. To optimize gas consumption, we save on Storage only simple data, keeping the large documents off-chain and storing here only their hash code and a link to retrieve them. The data related to the agri-food production events are stored in the Event-log, because they are not needed by the SCs of the system, but are only read by users' apps.
On average, this information storage choice reduces the gas consumption to 1/5, compared to using only the Storage.

Through a GUI, the system allows system actors to record events related to food production, and then a user to navigate through these events over time. 
All the concepts of the system are cataloged in a specific JSON file, with the relevant associated fields.
To make the system usable even for users with a low level of experience, these JSON files can be created through an easy-to-use back-office application.
In particular, we have a JSON files for:

\begin{itemize}

\item  The list of actors. For each actor id and name fields are specified.
\item  The list of the companies involved in the process (producer, processor or retailer). For each company name, resources (the list of ids of resources and products managed by that company), authorized actors (the list of ids of the actors authorized to record events for products related to that company) are specified.
\item For each type of productive resource or agro-industrial product managed by the production process, the fields id, type (R for resource or P for product), name, authorized actors (the list of ids of the actors authorized to record events for that product) are specified. Further fields can be automatically enhanced by the system when creating specific products: description, quantity (a positive integer describing the quantity of the product or the size of the productive resource), unit of measure (mandatory if quantity is used). 
\item The list of relevant events. 
Events are always associated with a product or a resource. For each event we specify id, name, products/resources (the list of products/resources ids that can be associated with that event), authorized actors (the list of ids of the actors authorized to record that event), type (D for documentation, T for transformation), generated products (the list of product ids that can be generated by that event), parameters (list of names and allowed types for the parameters associated to that event).
\item The list of activities that allows to customize the application template, based on the company and on the users.
\end{itemize}

Each of these files contains a key that determine the management of its content by the application.

\begin{figure}[h]
\centering\includegraphics[width=1\linewidth]{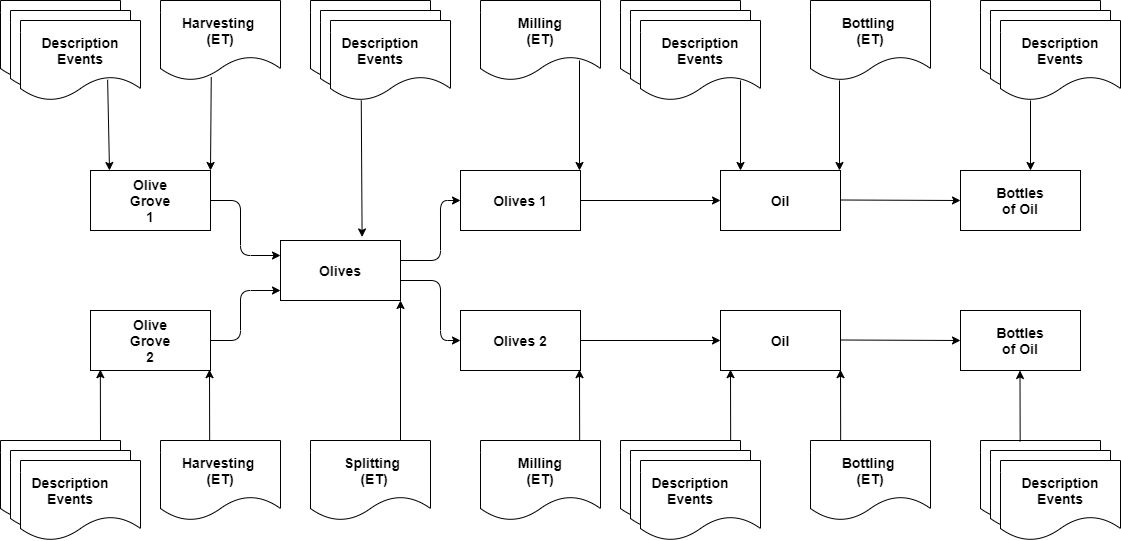}
\caption{Olive oil production flow.}\label{fig:OliveOilProduction}
\end{figure}

Figure \ref{fig:OliveOilProduction} shows the simplified olive oil production flow. The agro-industrial products are denoted as rectangles, while the associated events as "documents" in the flow chart notation (rectangles with wavy bottom side), possibly multiple.
Transformation events (denoted in the figure as ET), always single, are associated near the right side of products or resources, and cause them to be transformed into other products, or divided into batches. In the shown example, the olives come from two olive groves (1 and 2). Once harvested with a single event (Harvest) they are divided into two lots (Olive 1 and Olive 2), immediately before milling, which takes place separately for the two lots. Consequently, they have no other associated description events.
The flow of agro-industrial products, from left to right, can be represented by a tree: each product can have more "fathers", and more "sons". Consequently, the "navigation" backwards, starting from the last product, is not unique and must be managed in the user interface.

\section{conclusions and future works} \label{sec:Conclusions}

Traceability systems are considered important to ensure the safety of a food product and prevent food fraud in the food supply chain. 
System based on Blockchain and smart contracts for monitoring, integrated with the Internet of Things, allow a traceability system to occur without third party  which usually is responsible for controlling the agri-food supply chain,  transparency of data and indication of the origin of the products, as well as allowing the end customer to control the characteristics of the product purchased.

Specifically the advantages of the management system of an agri-food supply chain via blockchain, generated in a semi-automatic way, are:

\begin{itemize}
\item The consumer can be sure of the origin, the production process and the quality of the product purchased.
\item The Authority in charge of production control is facilitated in the controls, and can reduce on-site inspections.
\item The manufacturer can certify in a simple and non-falsifiable way all the steps of a production.
\item Development times and costs are reduced, while maintaining a high level of security and trust.
\item Blockchain system can also manage contractual transactions and payments, in this case, the system could also be extended to payments and the transfer of ownership of a product could also be marked by a cryptocurrency transfer.
\end{itemize}

To the best of our knowledge, this is the first attempt to automatically develop custom DApps for agri-food supply chain, by building configurable SCs to be assembled together and we truly believe in the innovation of the proposed method. Moreover, it was actually used to develop different real systems, thus confirming its efficiency.
Note that our flexible approach may lead to high gas cost in creating and updating SCs on Ethereum blockchain. A data represented as a string is way more costly than a native data, both in terms of storage and of computations needed to code and decode it. For this reason, we advise to use a permissioned blockchain publicly accessible for reading as a backbone. Future works may include the creation of a DSL (Domain Specific Language), the generalization to generic and non-ethereum-like blockchains, and also the generalization to other kind of supply chains.



\bibstyle{IEEEtran}
\bibliography{paper.bib}

\end{document}